\documentclass{ws-ijmpb}
\usepackage{graphicx}

\begin{document}

\markboth{Kh. P. Khamrakulov} {Regular and Chaotic Dynamics of a
Matter-Wave Soliton Near the Atomic Mirror}

\title{REGULAR AND CHAOTIC DYNAMICS OF A MATTER-WAVE SOLITON
NEAR THE ATOMIC MIRROR}

\author{KH. P. KHAMRAKULOV}
\address{Physical-Technical Institute, Uzbek Academy of Sciences, Tashkent, 100084, Uzbekistan \\
khamrakulov@uzsci.net}

\maketitle

\begin{history}
\received{Day Month Year} \revised{Day Month Year} \accepted{(Day
Month Year)} \comby{(xxxxxxxxxx)}
\end{history}

\begin{abstract}
The dynamics of the soliton in a self-attractive Bose-Einstein
condensate under the gravity are investigated. First, we apply the
inverse scattering method, which gives rise to equation of motion
for the center-of-mass coordinate of the soliton. We analyze the
amplitude-frequency characteristic for nonlinear resonance.
Applying the Krylov-Bogoliubov method for the small parameters the
dynamics of soliton on the phase plane are considered. Hamiltonian
chaos under the action of the gravity on the Poincar\'{e} map are
studied.
\end{abstract}


\section{Introduction}

Soliton is a localized nonlinear wave that propagate without
losing its shape due to equilibrium between dispersion and
nonlinearity effects.\cite{Scott} Solitons appear in such physical
systems as nonlinear optics, hydrodynamics and plasma waves.

The Bose-Einstein condensate (BEC) represents a giant matter-wave
packet. One of the most important aspects of matter-wave packets
is that they are strongly affected by gravity. In particular, they
fall towards earth like a beam of ordinary atoms. Since the
matter-wave packet is a superposition of macroscopic de Broglie
waves of ultracold massive atoms, they are accelerated under
gravity. This property of matter-waves was employed in the design
of an output coupler for the first atom laser,\cite{Mewes} and
demonstration of coherence of a freely expanding and overlapping
BEC.\cite{Andrews}

In physics, an atomic mirror is a media which reflects neutral
atoms in the similar way as the conventional mirror reflects
visible light. Atomic mirrors can be made of electric fields or
magnetic fields\cite{Merimeche} electromagnetic
waves\cite{Balykin} or just silicon wafer.\cite{Friedrich} In the
last case, atoms are reflected by the attracting tails, of the van
der Waals attraction (quantum reflection).\cite{Shimizu} Such
reflection is efficient when the normal component of the
wavenumber of the atoms is small or comparable to the effective
depth of the attraction potential. Roughly, the distance at which
the potential becomes comparable to the kinetic energy of the
atom. To reduce the normal component most atomic mirrors are
blazed at the grazing incidence. At grazing incidence, the
efficiency of the quantum reflection can be enhanced by a surface
covered with ridges.\cite{Fujita,Kouznetsov}

Recently quantum reflection of matter-waves from a solid surface
has been the subject of considerable interest both from the
viewpoints of basic physics and BEC applications. Specifically,
matter-wave dynamics near the solid surface can be a very
sensitive probe for the Casimir force.\cite{Pasquini} Meantime,
atom chips, where a BEC is stored and manipulated near the solid
substrate, open up new perspectives for application.\cite{Ott}
Coherent acceleration of matter-wave packet falling under gravity
and bouncing off a modulated magnetic mirror showed the
possibility to realize the Fermi acceleration with
matter-waves.\cite{Saif}

This work is aimed at investigation of the dynamics of a
matter-wave soliton near the solid surface under the action of a
linear potential, originating from the attractive force of
gravity. The effect of a solid surface is modelled by a reflecting
delta-potential barrier. In real experiments such a barrier can be
created by means of a laser light far-off blue-detuned from atomic
transitions. The underlying mathematical model is based on the one
dimensional Gross-Pitaevskii equation (GPE) for the BEC with a
negative atomic scattering length, when the GPE supports
self-localized solution, the so called matter-wave soliton.

The paper is organized as follows. In Sec. 2, we describe
interactions between the matter-wave soliton and delta-potential
barrier. In Sec. 3, the Krylov-Bogoliubov method applied to the
equation of motion for the center-of-mass coordinate of the
soliton. In Sec. 4, we consider the Poincar\'{e} map for nonlinear
resonance. In concluding Sec. 5, we summarize our results.

\section{The Model and Governing Equation}

The dynamics of a BEC in the mean-field approximation at zero
temperature is governed by the 3D GPE\cite{Pethick,Pitaevskii}
\begin{equation}
i\hbar\frac{\partial\Psi}{\partial
t}=\left[-\frac{\hbar^2}{2m}\nabla^{2}+V(\mathbf{r})+\frac{4\pi
\hbar^2\alpha_{s}N}{m}|\Psi|^{2}\right]\Psi, \label{gpe1}
\end{equation}
where $\Psi(\mathbf{r},t)$ is the macroscopic wave function of the
condensate normalized so that $\int^{\infty}_{-\infty}
|\Psi(\mathbf{r},t)|^{2}d\mathbf{r}=1$, $N$ is the total number of
atoms, $m$ is the atomic mass, $\alpha_{s}$ is the $s-$wave
scattering length (below we shall be concerned with an attractive
BEC for which $\alpha_{s}<0$), and
\begin{equation}
V(\mathbf{r})=\frac{m}{2}\big[\omega^{2}_{x}x^2+\omega^{2}_{\perp}
(y^2+z^2)\big] \label{trap}
\end{equation}
is the axially symmetric trapping potential which provides for
tight-fitting confinement in the transverse plane $(y,z)$, as
compared to free axial trapping, assuming $\omega^{2}_{x}/
\omega^{2}_{\perp}\ll 1$.

When the transverse confinement is strong enough, so that the
transverse oscillation quantum $\hbar\omega_{\perp}$, is much
greater than the characteristic mean-field interaction energy
$N|\alpha_{s}||\Psi|^{2}$, the dynamics is effectively one
dimensional. In this case, the 3D wave function may be effectively
factorized as $\Psi(x,y,z,t)=\psi(x,t)\varphi(y,z)$, where
$\varphi(y,z)=\exp\big[-(y^{2}+z^{2})/2l_{\perp}^{2}\big]/
\sqrt{\pi}l_{\perp}$ is the normalized ground state of the 2D
harmonic oscillator in the transverse direction, with
$l_{\perp}=\sqrt{\hbar/m\omega_{\perp}}$ being the corresponding
transverse harmonic oscillator length. Substituting the factorized
expression into the 3D GPE (\ref{gpe1}), and integrating it over
the transverse plane (y,z), one derives the effective 1D GPE for
an attractive BEC
\begin{equation}
i\hbar\frac{\partial\psi}{\partial
t}=\left[-\frac{\hbar^2}{2m}\frac{\partial^2}{\partial
x^2}+V(x)-q_{1D}|\psi|^2\right]\psi, \label{gpe2}
\end{equation}
where we have neglected the zero-point energy of the transverse
motion $\hbar\omega_{\perp}$, and defined the coefficient of the
1D nonlinearity, $q_{1D}=4\pi|\alpha_{s}|\hbar\omega_{\perp}
\int^{\infty}_{-\infty}|\varphi(y,z)|^{4}dydz=2|\alpha_{s}|
\hbar\omega_{\perp}$ and $V(x)=m\omega^{2}_{x}x^2/2$ is the axial
parabolic trap in the x direction.

Let us consider the case when the BEC falls under gravity force
and bouncing off from the modulated atomic mirror. Next, we shall
assume that the axially parabolic trap in the Eq. (\ref{gpe2}) can
be changed by the linear potential and delta-potential barrier. As
a result, the 1D GPE, taking into account the gravity can be
written in the following form:
\begin{equation}
i\hbar\frac{\partial\psi}{\partial
t}=\left[-\frac{\hbar^2}{2m}\frac{\partial^2}{\partial
x^2}+V(x,t)-q_{1D}|\psi|^2\right]\psi. \label{gpe3}
\end{equation}
The potential $V(x,t)$ for the 1D GPE (\ref{gpe3}) with the
falling BEC under the gravity has the following form:
\begin{eqnarray}
V(x,t) & = & V_{1}(x)+V_{2}(x,t), \\
V_{1}(x) & = & k x, \\
V_{2}(x,t) & = & V_{0}\delta\left[x-f(t)\right], \label{deltapot}
\end{eqnarray}
where $V_{1}(x)$ is the lineal potential, $k=mg$, $g$ is the
acceleration of gravity, $V_{2}(x,t)$ is the delta-potential
barrier whose position is oscillating with the amplitude of
external force $f_{0}$ and time dependence function given by
$f(t)=f_{0}\sin(\gamma t+\phi)$, $\gamma$ and $\phi$ are the
frequency and the phase of the amplitude of external force.

For the purposes of further simplification, let us rewrite Eq.
(\ref{gpe3}) using dimensionless variables: $t\rightarrow
t\omega_{\bot}/2, x\rightarrow x/l_{\bot}, l_{\bot}=
\sqrt{\hbar/m\omega_{\bot}}, l\rightarrow
\left(l_{\bot}/l_{g}\right), l_{g}^{-3}=2m^{2}g/\hbar^2,
V_{0}\rightarrow 2V_{0}/(\hbar\omega_{\bot}l_{\bot})$, and the
rescaled wave function $u\rightarrow \sqrt{2|\alpha_{s}|}\psi$,
\begin{equation}\label{dimgpe}
iu_{t}+u_{xx}+V(x,t)u+2|u|^2u=0.
\end{equation}
It is well known, in the absence of the potential term $V(x,t)=0$,
Eq. (\ref{dimgpe}) gives rise to a commonly known family of
soliton solutions,\cite{Zakharov}
\begin{eqnarray}
u(x,t) & = & 2i\eta\frac{\exp\left[-2i\xi x-4i(\xi^2-\eta^2)t-
i\phi_{0}\right]}{\cosh[2\eta(x-\zeta)]}, \\
\zeta & = & -4\xi t+\zeta_{0}, \label{exactsol}
\end{eqnarray}
where $\eta,\xi,\zeta$ are, respectively, the amplitude, velocity,
center-of-mass coordinate and $\zeta_{0},\phi_{0}$ are the initial
coordinate and phase.

If we can consider the effects of the linear potential $V_{1}(x)$
and delta-potential barrier $V_{2}(x,t)$ as perturbations for the
soliton
\begin{equation}
iu_{t}+u_{xx}+2|u|^{2}u=\epsilon R, \label{perturb}
\end{equation}
\begin{equation}
\epsilon R=\left[V_{1}(x)+V_{2}(x,t)\right]u.
\end{equation}

\begin{figure}[htb]
\centerline{\includegraphics[width=8cm,height=6cm,clip]{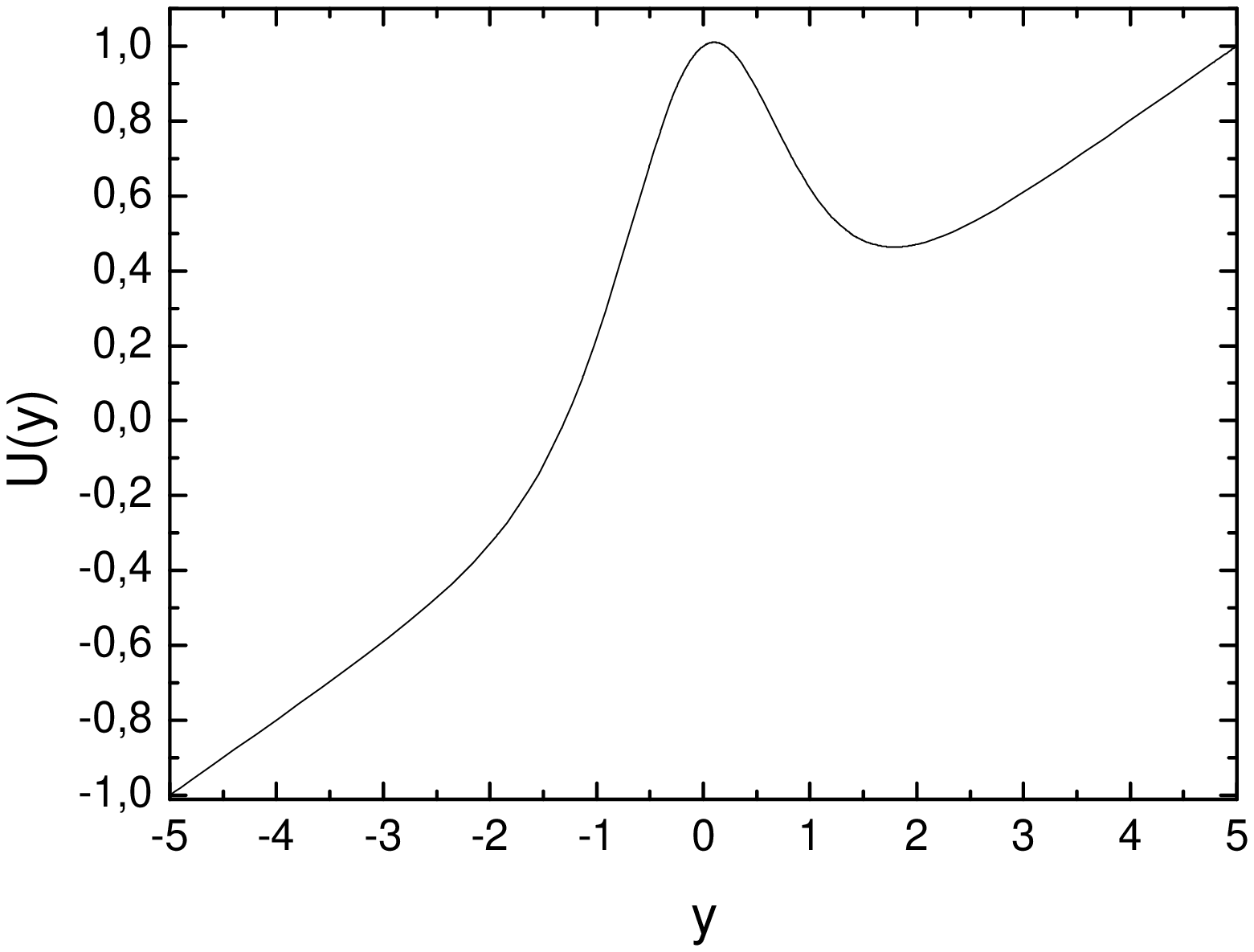}}
\caption{The shape of the anharmonic potential given by Eq.
(\ref{pot}) for the next parameter values: $k=-0.1$, $V_{0}=-1$,
$\eta=0.5$, $y_{0}=1.78$.} \label{fig1}
\end{figure}

Applying the conservation law of the field momentum $dP/dt=0$ from
the soliton theory\cite{Zakharov} and taking into account the
$dP/dt=i\int_{-\infty}^{\infty}(u_{t}u_{x}^{\ast}-u_{x}u_{t}^{\ast})dx$,
we finally get the following equation for the soliton
center-of-mass coordinate, which has the following form:
\begin{equation}
\frac{d^{2}\zeta}{dt^{2}}=-2k-8V_{0}\eta^{2}\frac{\sinh\left[2\eta
(\zeta-f(t))\right]}{\cosh^3\left[2\eta(\zeta-f(t))\right]}.
\label{goveq1}
\end{equation}
Introducing the new variable $y=\zeta-f(t)$, one obtains from Eq.
(\ref{goveq1}) the following equation
\begin{equation}
\frac{d^{2}y}{dt^2}=-\frac{\partial U}{\partial y}-\ddot f(t).
\label{goveq2}
\end{equation}
This is the governing equation for a unit mass quasi-particle
moving in the field of anharmonic potential (Fig. \ref{fig1})
\begin{equation}
U(y)=2ky-\frac{2V_{0}\eta}{\cosh^{2}(2\eta y)}, \label{pot}
\end{equation}
and external force $F(t)=-\ddot f(t)=f_{0}\gamma^{2} \sin(\gamma t
+\phi)$.

Let us expand the potential $U(y)$ by a series, according to the
$x=y-y_{0}$ degree of deviation from the point of equilibrium,
with fourth order of approximation inclusively, then, the equation
of motion (\ref{goveq2}) can be written as follows:
\begin{equation}
\ddot x+\omega^2x+\alpha x^{2}+\beta
x^{3}=f_{0}\gamma^{2}\sin(\gamma t+\phi), \label{lineq}
\end{equation}
with the coefficients
\begin{eqnarray}
\omega^2 & = & 16V_{0}\eta^{3}{\rm sech^2}\left(2\eta
y_{0}\right)\left[3{\rm sech}^2\left(2\eta y_{0}\right)-2\right],
\\
\alpha & = & 64V_{0}\eta^{4}\tanh\left(2\eta y_{0}\right){\rm
sech^2}\left(2\eta y_{0}\right)\left[1-3{\rm sech^2}\left(2\eta
y_{0}\right)\right], \\
\beta & = & (32/3)V_{0}\eta^{5}{\rm sech}^6\left(2\eta
y_{0}\right)\left[26\cosh(4\eta y_{0})-\cosh(8\eta
y_{0})-33\right], \label{freq}
\end{eqnarray}
where $\omega$ is the frequency of the quasi-particle, $\alpha$
and $\beta$ are the anharminic coefficients, $\gamma$ is the
frequency of the external force and $\phi$ is the phase.
\begin{figure}[htb]
\centerline{\includegraphics[width=6cm,height=5cm]{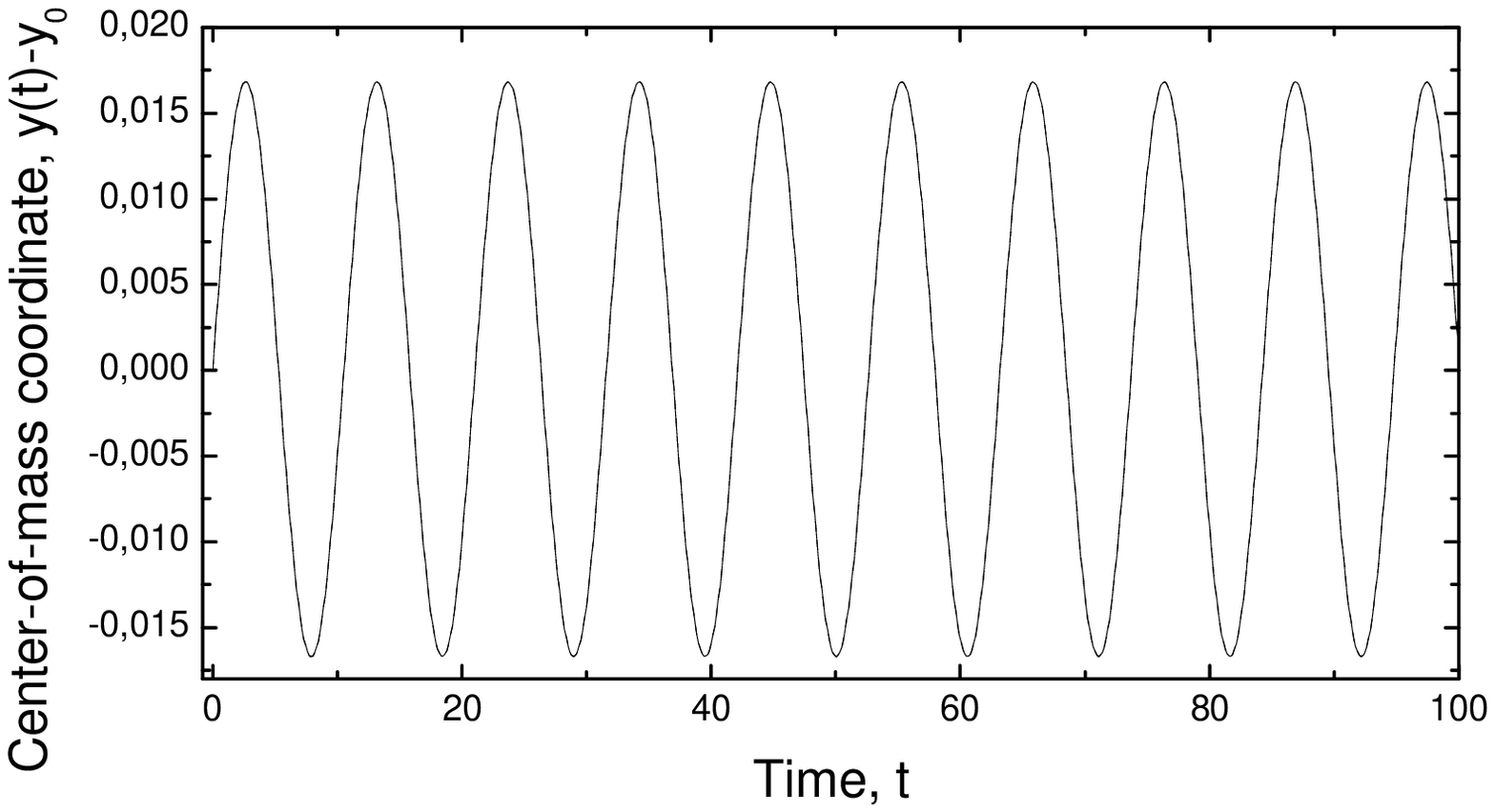}\quad
             \includegraphics[width=6cm,height=5cm]{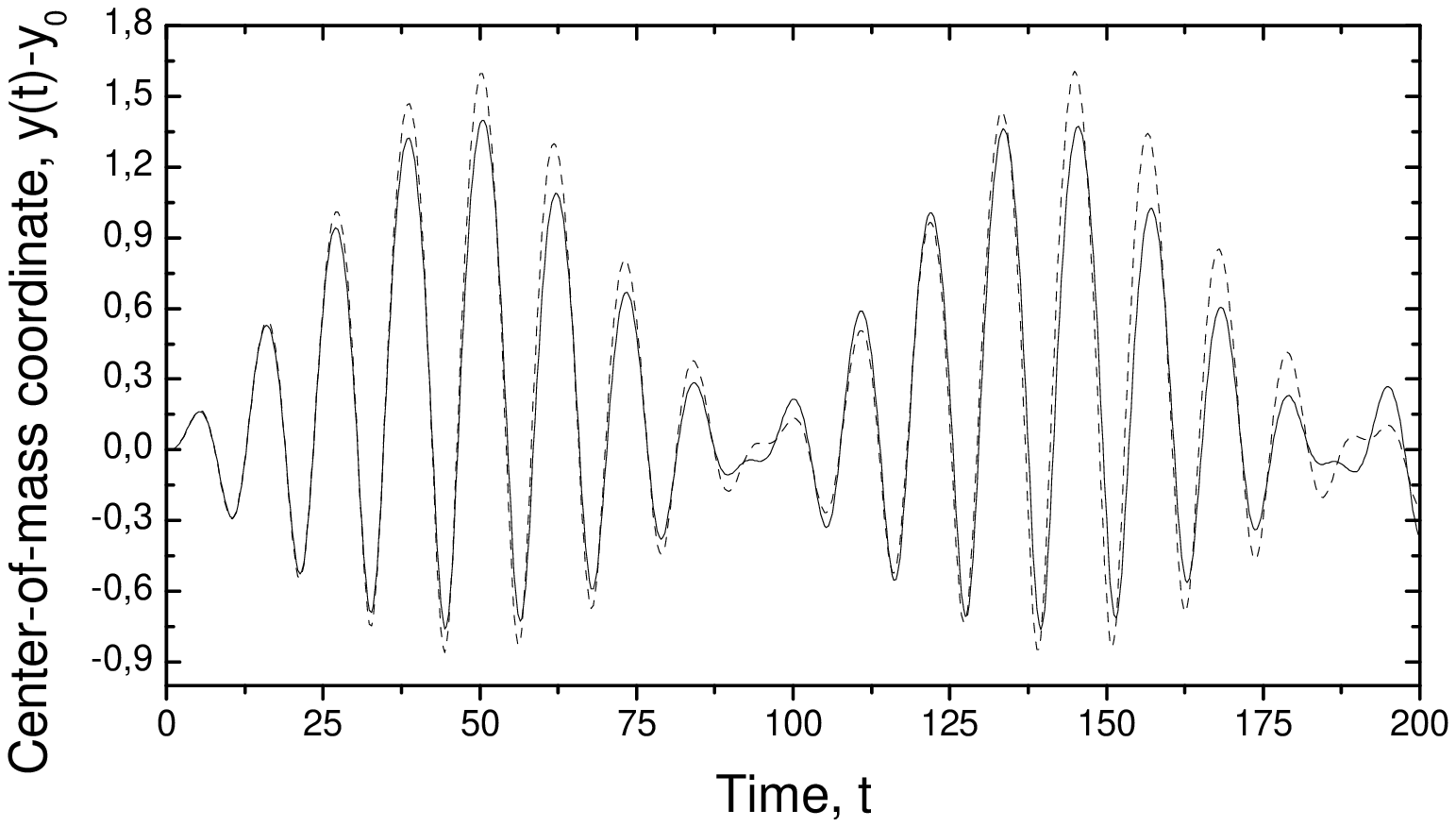}}
\caption{The dynamics of the center-of-mass coordinate of soliton.
Numerical simulation of GPE (\ref{dimgpe}) (solid line) and
governing equation (\ref{goveq2}) (dashed line). Left panel: The
case when the delta-potential barrier do not oscillating.
Parameters are: $V_{0}=-1, \eta=0.5, y_{0}=1.78, \omega=0.596,
\alpha=-0.273, \beta=0.081, a=0.39, f_{0}=0$. Right panel:
Delta-potential barrier as perturbation of the external force.
Parameter values: $\gamma=\omega$, $f_{0}=0.1$ (beats).}
\label{fig2}
\end{figure}

It is interesting to consider the dynamics of the interaction
between the soliton and the oscillating surface in the proximity
of the resonance, which is illustrated in the Fig. \ref{fig2}. The
left panel of Fig. \ref{fig2}, illustrates the cases when the
interacting surface does not oscillating, the soliton oscillates
harmonically with the period $T=2\pi/\omega\left[ 1+(5a^2/12
\omega^4-3\beta/8\omega^2)a^2+o(a^2) \right]$, as described in
Ref. 18. Thus, near a position of stable equilibrium, a system
executes harmonic oscillations. As can be seen in the right panel
of Fig. \ref{fig2}, the dependence of the amplitude $a$ of the
forced oscillations on the frequency of the external force has the
characteristic resonance shape: The nearer the frequency of the
external force to the natural frequency $\omega$, the more the
external force rocks the system. The phase $\phi$ of the forced
oscillations undergoes a jump of $-\pi$ as $\gamma$ passes thought
the resonance frequency $\omega$. When $\gamma$ is near $\omega$,
beats are observed, i. e., the amplitude of the quasi-particle
alternately waxes, when the relation of the phases of the
quasi-particle and the external force is such that the external
force rocks the quasi-particle, communicating energy to it and
wanes, when the relation between the phases changes in such a way
that the external force brakes the quasi-particle. The closer
frequencies $\gamma$ and $\omega$, the more slowly the phase
relation changes and the larger the period of the beats. As
$\gamma\rightarrow\omega$, the period of the beats approaches
infinity.

In order to estimate the actual values of the time and space
units, we shall provide the experimental parameters from Ref. 16.
The current experiment considers a single soliton of the lithium
condensate. The $s-$wave scattering length, at the value of the
magnetic field $B=425$ $\mathrm{G}$ (which was used to make the
atomic interaction attractive, via Feshbach resonance), was
$a_{s}=-0.21$ $\mathrm{nm}$. With the mass of a $^{7}\mathrm{Li}$
atom, $m=11.65\times10^{-27}$ $\mathrm{kg}$, we have the following
time and space units: $\omega_{x}^{-1}\simeq3\times10^{-3}$
$\mathrm{s}$, $l_{\perp}=\sqrt{\hbar/m\omega_{\perp}}\simeq12$
$\mu m$, the trap's aspect ratio being
$\omega_{x}/\omega_{\perp}\simeq7\times10^{-2}$.

\section{The Dynamics in the Potential Field}

The case of a small nonlinearity turns out to be more complicated,
strange though it may seem, if, perhaps, a more interesting one,
as is demonstrated by a curious example in Ref. 17. When the
anharmonic terms in forced oscillations of a system are taken into
account, the phenomena of resonance acquire new properties. Let
$\gamma=\omega+\Delta$, with small $\Delta$, i. e. $\gamma$ be the
resonance value. Strictly speaking, when nonlinear terms are
included in the equation of the free oscillations, the term higher
order in the amplitude of external force (such as occur if it
depends on the displacement $x$) should also be included. We shall
omit these terms merely to simplify the formulae, i. e. they do
not affect the qualitative results. As well known,\cite{Stoker} in
the linear approximation, the amplitude $a$ is given near
resonance, as a function of the amplitude $f_{0}$ and frequency
$\gamma$ of the external force, which we write as
$a^{2}\varepsilon^{2}=f_{0}^{2}/4\omega^{2}$. The nonlinearity of
the oscillations results in the appearance of an amplitude
dependence of the eigenfrequency, which we write as
$\omega+(3\beta/8\omega)a^{2}$. Accordingly, we replace $\omega$
by $\omega+(3\beta/8\omega)a^{2}$ (or, more precisely, in the
small difference $\gamma-\omega$). With $\Delta=\gamma-\omega$,
the resulting equations is
\begin{equation}
a^{2}\left(\Delta-\frac{3\beta}{8\omega}a^{2}\right)^{2}=
\frac{f_{0}^{2}}{4\omega^{2}}. \label{ampl}
\end{equation}

\begin{figure}[htb]
\centerline{\includegraphics[width=8cm,height=6cm,clip]{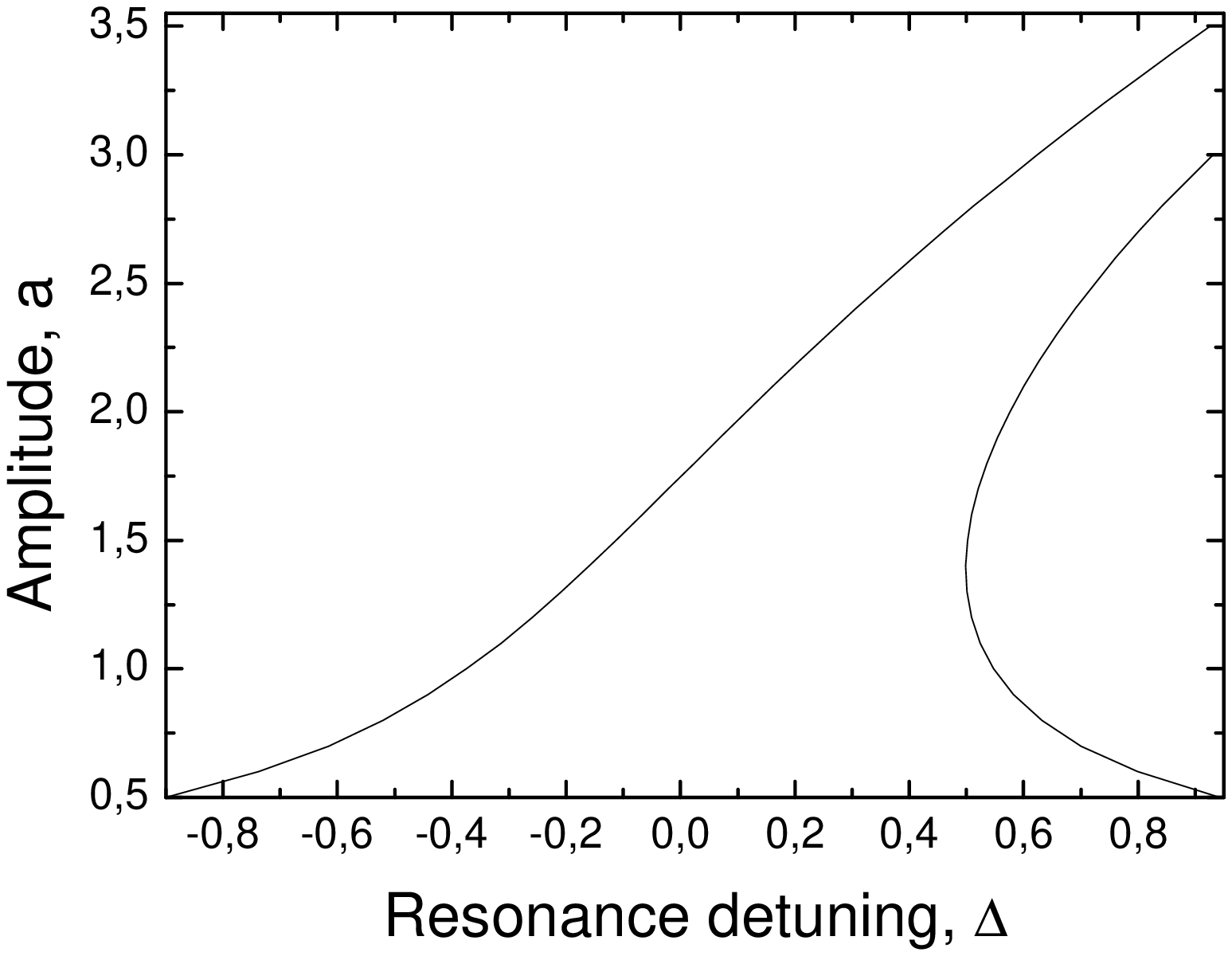}}
\caption{Amplitude-frequency characteristic for nonlinear
resonance at $\omega=0.596, \beta=0.081, f_{0}=0.55$.}
\label{fig3}
\end{figure}

Eq. (\ref{ampl}) is a cubic equation in $a^{2}$, and its real
roots give the amplitude of the external forced oscillations. Let
us consider how this amplitude depends on the frequency of the
external force for a given amplitude $f_{0}$ of that force. As
$f_{0}$ increases, the curve changes its shape, though at first it
retains its single maximum, which moves to positive $\Delta$ if
$3\beta/8\omega>0$. At this stage only one of the three roots of
Eq. (\ref{ampl}) is real. When $f_{0}$ reaches a certain value
$f_{\mathrm{th}}$ (to be determined below), however, the nature of
the curve changes. For all $f_{0}>f_{\mathrm{th}}$ there is a
range of frequencies in which Eq. (\ref{ampl}) has three real
roots. In the absence of friction, the damping coefficient is
zero. Consequently the Fig. \ref{fig3} indicates that in our case
the damping coefficient, which affects the knee of the curve, is
zero leading to the branches of the amplitude-frequency
characteristic receding into infinity.

It is widely known that nonlinear oscillating systems with weakly
nonlinearity can be studied by the perturbation theory methods.
Let us consider the dynamics of the soliton in one dimension, and
writing the Eq. (\ref{lineq}) in the form:
\begin{equation}
\ddot x+x=\varepsilon Q(x,\nu\tau), \label{perteq}
\end{equation}
where $\varepsilon Q(x,\nu\tau)=-\delta x^{2}-\lambda x^{3}+
\chi\sin(\nu\tau)$ is the periodic function with respect to
$\nu\tau$ with period $2\pi$, $\dot x=dx/d\tau, \tau=\omega t,
\delta=\alpha/\omega^2, \lambda= \beta/\omega^2, \chi=f_{0}\nu^2,
\nu=\gamma/\omega$ are the dimensionless variables, $\varepsilon$
is the small positive parameter, indicating the smallness of the
function $\varepsilon Q(x,\nu\tau)$ with regard to the linear term
(the order of smallness of the terms in this equation is
determined, so that with $\varepsilon\rightarrow0$ there is the
case for linear harmonic oscillations). In this case, we are
considering the main resonance, i. e. $\nu=1$. Higher order
resonance appear when $\omega\approx(n/r) \gamma$, where $n$ and
$r$ are integers. Taking into account the smallness of nonlinear
terms $(x\ll 1)$ with regard to the linear terms for solving Eq.
(\ref{perteq}) in zero-order approximation $x(\tau)$ can be chosen
as\cite{Bogoliubov}:
\begin{equation}
x(\tau)=b(\tau)\cos\left[\sigma(\tau)\right], \label{zeroapp}
\end{equation}
where $\sigma(\tau)=\nu\tau+\theta(\tau)$ and $b$, $\theta$ are
slowly varying functions of $\tau$. Using the Krylov-Bogoliubov
method (KBM) for the small parameters\cite{Bogoliubov} it is
possible to obtain the following coupled equations for $b$ and
$\theta$:
\begin{eqnarray}
\frac{db}{d\tau} & = & -\frac{\chi}{2\nu}\cos\theta, \\
\frac{d\theta}{d\tau} & = & \frac{\rho}{2\nu}+\frac{3\mu
b^2}{8\nu}+\frac{\chi}{2\nu b}\sin\theta, \label{coupeq}
\end{eqnarray}
where $\rho=1-\nu^2, \rho<<1, \mu=\lambda\left[1-10\delta^2/
(9\lambda)\right]$.

\begin{figure}[htb]
\centerline{\includegraphics[width=8cm,height=6cm,clip]{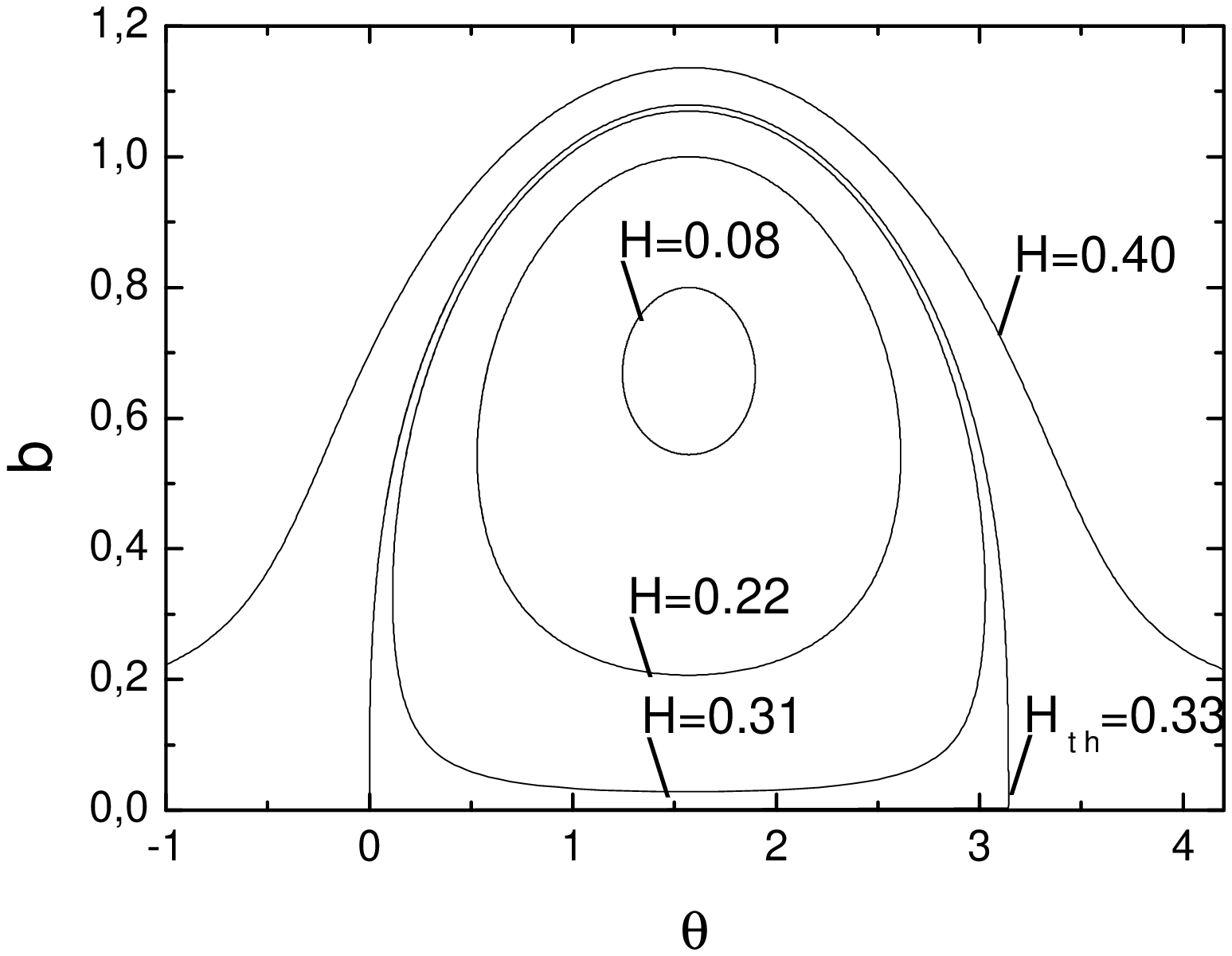}}
\caption{The phase trajectories for system (23)-(\ref{coupeq}).
The curve for $H_{\mathrm{th}}=0.33$ corresponds to the
separatrix; the curve for $H=0.4$ corresponds to the drift
trajectory. Parameters are: $\chi=0.1, \nu=1, \rho=0,
\mu=-0.423$.} \label{fig4}
\end{figure}

The equations (23)-(\ref{coupeq}) can be transformed into
Hamiltonian form by changing the new variable $b=\sqrt{p}$. The
Hamilton's equations are $dp/dt=-\partial H/\partial\theta,
d\theta/dt=\partial H/\partial p$, with Hamiltonian
$H=(\chi\sqrt{p}/\nu)\sin\theta+\rho p/(2\nu)+3\mu p^{2}/(16\nu)$.
Hamiltonian is conserved quantity, i. e. $H$ is the integral of
motion for system (23)-(\ref{coupeq}). Phase trajectories
corresponding to different values of $H$ at fixed $\chi$ and
$\rho$ are shown in Fig. \ref{fig4}. From qualitative analysis of
the system (23)-(\ref{coupeq}) it can be revealed that there
exists a separatrix $H=0$, which separates finite trajectories
corresponding to nonlinear resonances from infinite ones. Figure
\ref{fig4} illustrates the quasi-particle motion on the phase
plane at threshold Hamiltonian $H_{\mathrm{th}}= 0.33$ corresponds
to the separatrix. When the value of Hamiltonian is above
threshold, i. e. $H=0.4$, then we observe the quasi-particle can
leave the potential well, performing nonlinear oscillations.

\section{The Dynamics on the Poincar\'{e} Map}

In this section, we will describe the criterion of chaotic
oscillations in the problem of quasi-particle motion in the
potential field $U(y)$ under the action of periodic force $F(t)$
(see section 1). The force, which leads to such kind of the
motion, is potential. It is known,\cite{Naifeh,Tsoy} that after a
transient process there is a steady state in the system
(\ref{lineq}), i. e. $x(t)\sim a\cos(\gamma t+\vartheta_{0})$,
where $a$ and $\vartheta_{0}$ are amplitude and initial phase of
oscillations respectively. One can obtain by Lindstedt
method\cite{Stoker,Moon} the relation between
$\alpha,\beta,\gamma,\omega$ and amplitude $a$:
$F(\alpha,\beta,\gamma,\omega,a)=f_{0},$ where $F$ is some
function. This relation may be considered as an equation for
threshold amplitude $f_{0\mathrm{th}}$. As such, substituting this
solution to Eq. (\ref{lineq}) and equating the coefficients of
trigonometric functions we finally obtain the following condition
for breaking of the bound state:
\begin{equation}
f_{0}\geq f_{0\mathrm{th}}=a_{\mathrm{th}} \bigg\{\left[\left
(\omega/\gamma\right)^2-1+ 3\beta a_\mathrm{th}^2/
4\gamma^2\right]^2+(3/4)\left(\alpha a_{\mathrm{th}}/\omega
\gamma^2\right)^2\bigg\}.
\end{equation}

This criterion is determine the transition boundary from periodic
motion to the chaotic one. It is important to draw a line between
the systems of the damped oscillations and the ones without such.
In the systems without the damped oscillations or weakly damped
oscillations the Poincar\'{e} map of the chaotic motion often has
a form of disordered clusters of points on the Poincar\'{e} map.
Such motions are called stochastic. Under the action of external
force
\begin{figure}[htb]
\centerline{\includegraphics[width=6cm,height=5cm]{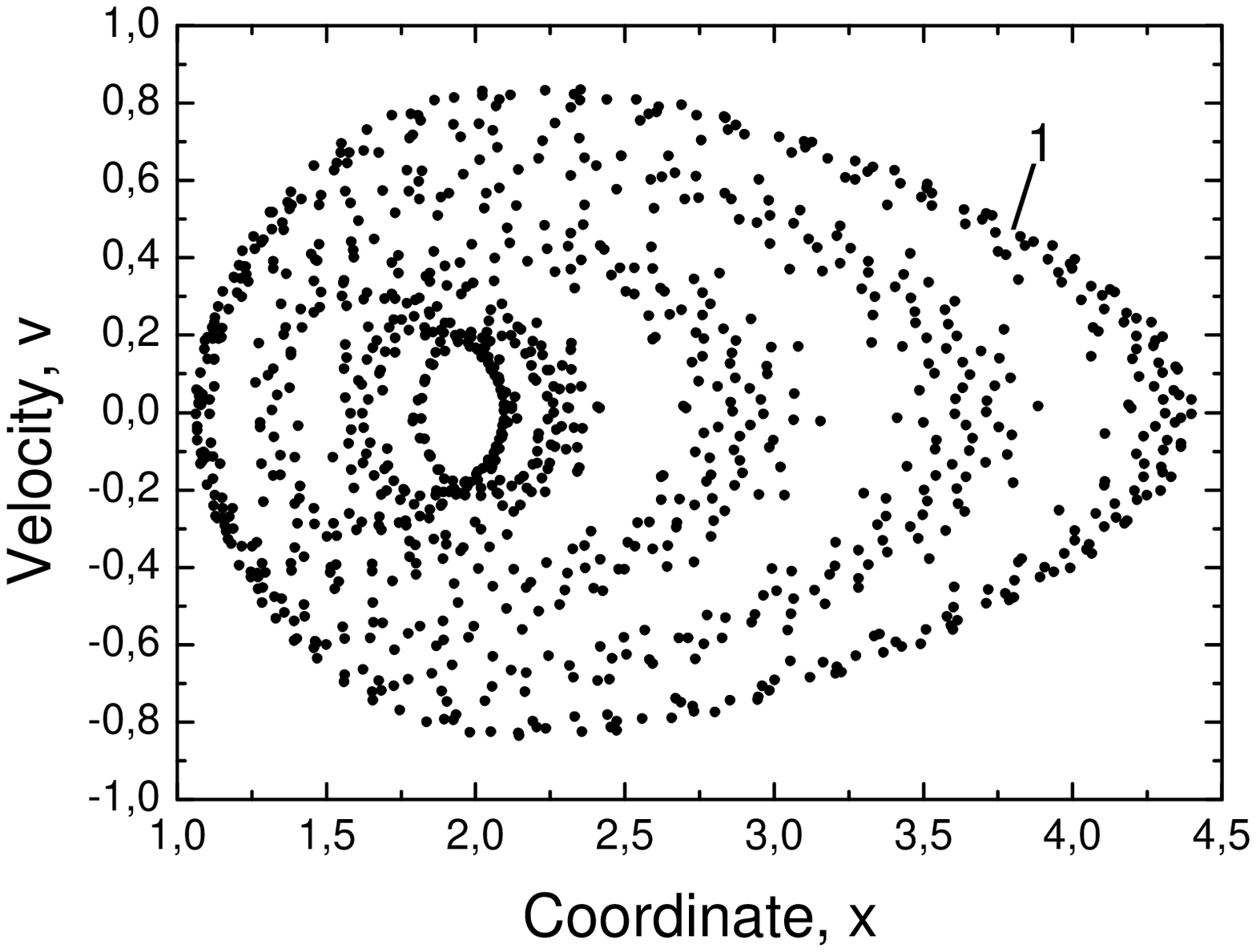}\quad
            \includegraphics[width=6cm,height=5cm]{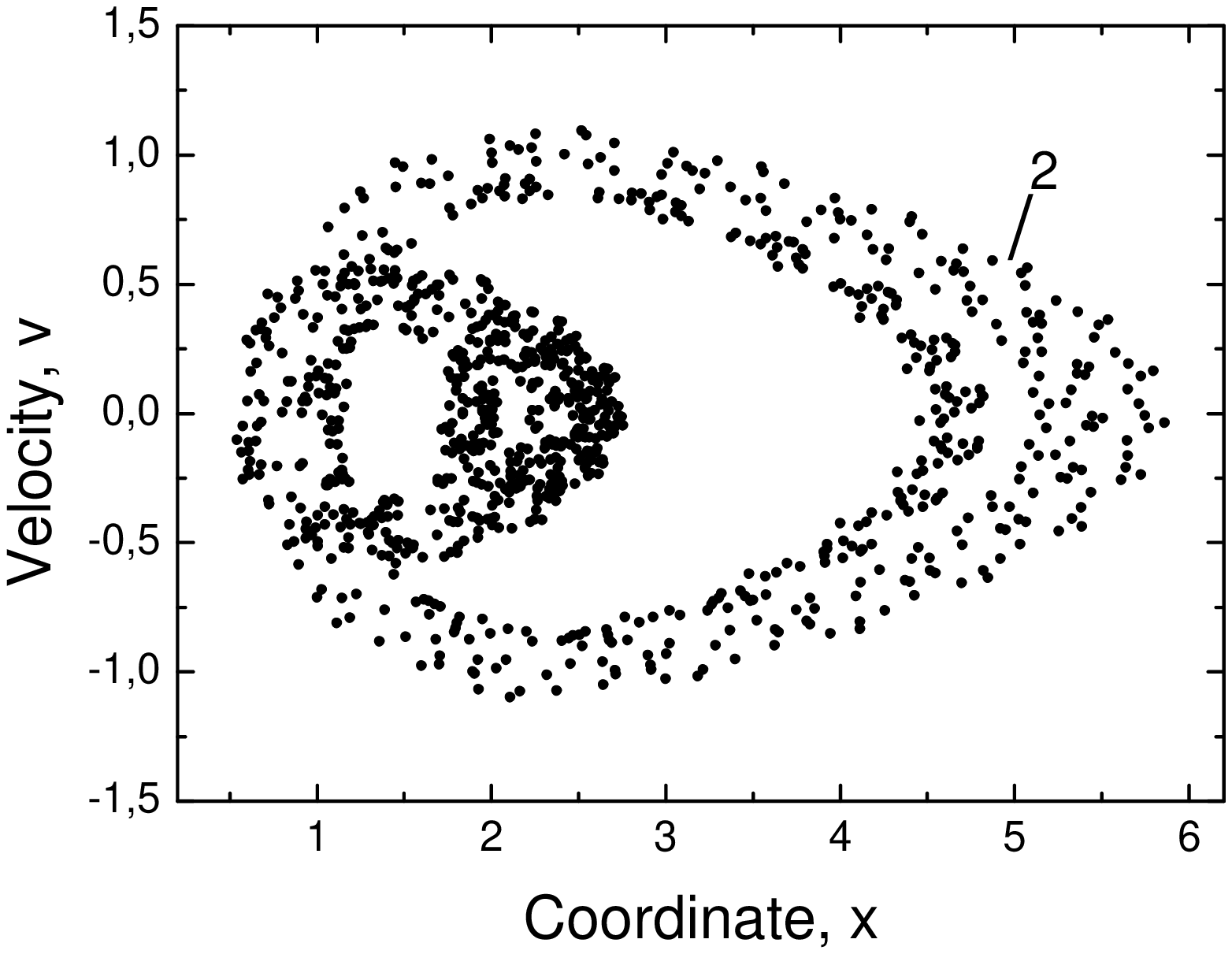}}
\caption{The Poincar\'{e} map of Eq. (\ref{lineq}). Left panel:
The set of points 1 corresponds to the quasi-periodic motion with
the next parameters: $\omega=0.596, \alpha=-0.273, \beta=0.081,
f_{0}=0.45$. Right panel: The set of points 2 corresponds to the
chaotic motion and going away to $+\infty$ with
$f_\mathrm{0th}=0.55$.} \label{fig5}
\end{figure}
near trajectories with the period $T=nT_{\mathrm{ext}}$, where
$T_{\mathrm{ext}}=2\pi/\omega$ and $n$ is integer, nonlinear
resonance occurs.\cite{Chirikov} Figure \ref{fig5} illustrates the
Poincar\'{e} map, as can be seen in the left panel, the set of
points 1 represents quasi-periodical motion, the separatrix and
stationary points of the Poincar\'{e} map appear due to nonlinear
resonance. The width of the separatrix of the nonlinear resonance
increases with growing of $f_{0}$, and at some values of $f_{0}$
the overlap of neighboring resonances occurs. In Fig. \ref{fig5},
we show that if the amplitude of the external force changes from
$f_{0}=0.45$ to $f_{0}=0.5408$, then the quasi-periodic motion of
the quasi-particle occurs. As can be seen in the right panel of
Fig. \ref{fig5}, the chaotic motion of the quasi-particle occurs
as the value of the amplitude of the external force approaches the
threshold value of $f_{0\mathrm{th}}=0.55$. This threshold
amplitude separates the quasi-periodic motion from the chaotic
motion of the quasi-particle. The set of points 2 represents
chaotic motion of the quasi-particle. As a result, the motion of
the quasi-particle with the threshold amplitude on the
Poincar\'{e} map becomes random, i. e. Hamiltonian chaos appears
in the system.\cite{Chirikov,Lichtenberg}

\begin{figure}[htb]
\centerline{\includegraphics[width=8cm,height=6cm,clip]{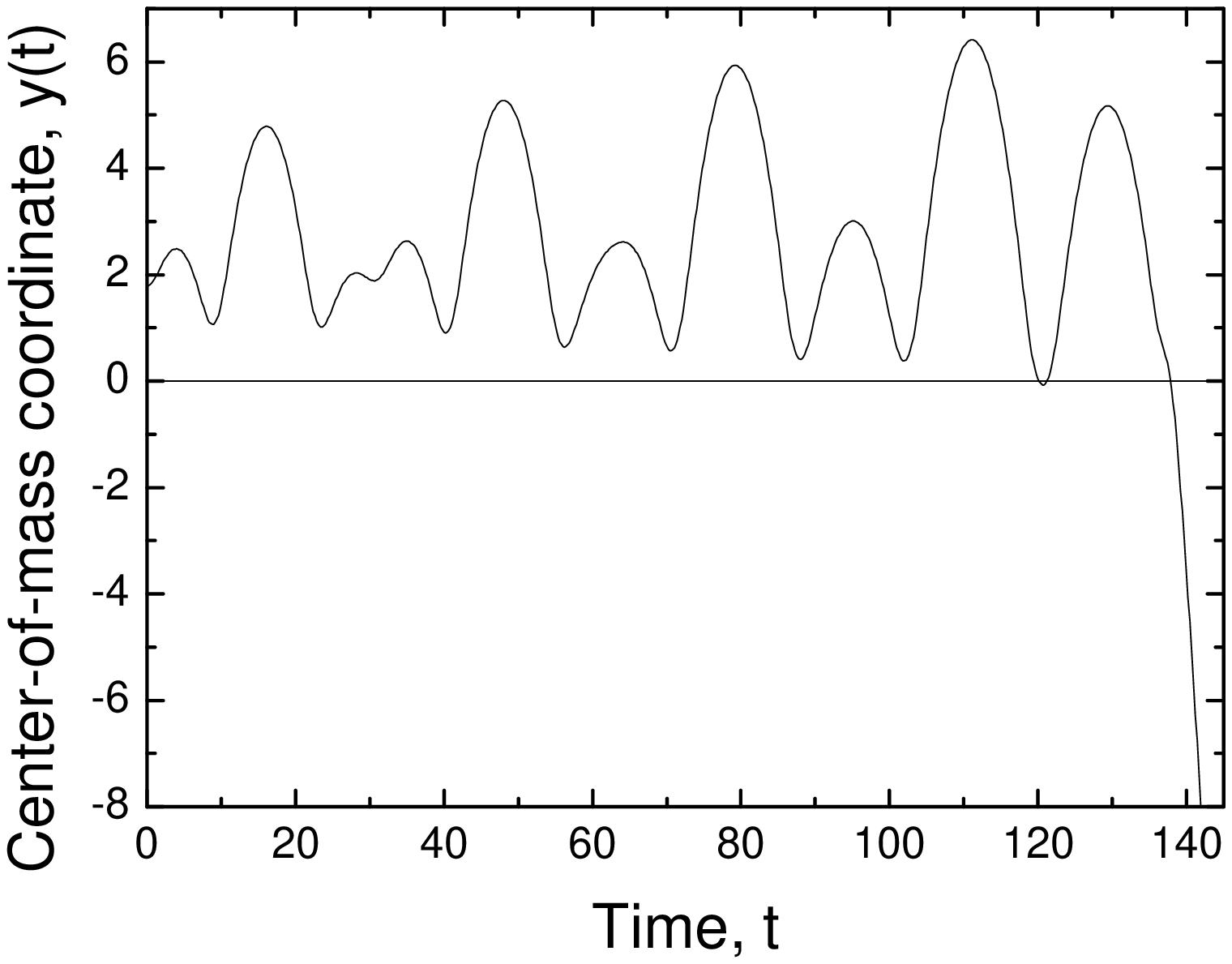}}
\caption{The result of numerical integration of Eq. (\ref{goveq2})
for the next parameters: $V_{0}=-1, \eta=0.5, y_{0}=1.788,
\omega=0.596, \alpha=-0.273, \beta=0.081, f_\mathrm{0th}=0.55$.}
\label{fig6}
\end{figure}

The typical picture of escape of the quasi-particle from the
potential well under the action of resonance force is shown in the
Fig. \ref{fig6}. On the other hand, the quasi-particle leaves the
potential well when its kinetic energy becomes comparable with the
value corresponding to the difference between the potential well
bottom and the separatrix. When the amplitude of the external
force reaches its threshold, the quasi-particle accelerates to the
boundary to leave the potential well. Figure \ref{fig6}
illustrates the mean-first passage time\cite{Hanggi} (MFPT) of the
quasi-particle, as a result of numerical integration of governing
equation (\ref{goveq2}) at $f_\mathrm{0th}=0.55$.

\section{Conclusion}

We have studied the nonlinear effect the matter-wave soliton
interacting with the delta-potential barrier by means of nonlinear
mechanics and numerical simulations. The Krylov-Bogoliubov method
provides a framework to understand the dynamics of the
quasi-particle on the phase plane. Effects of nonlinear
resonances, are studied by perturbation theory. By applying the
Poincar\'{e} map, we showed as the amplitude of the external force
approaches to the threshold value, a Hamiltonian chaos can be
observed in the system. The developed model predicts that the
quasi-particle, being on the bottom of the potential well, begins
to scatter through resonances and escapes the potential well,
increasing stochastically its energy. The results can be useful in
development of new methods aimed at the delta-potential barrier by
scattering solitons on them.

Two related physical phenomena have recently been observed:
quantum states of ultra-cold neutrons in the gravitational field
above a flat mirror, and quantum states of cold neutrons in an
effective centrifugal potential in the vicinity of a concave
mirror.\cite{Nesvizhevsky} Obtained results can be useful for
experiments with ultra-cold neutrons and ultra-cold quantum gases.

\section*{Acknowledgements}

Author grateful to thanks F. Kh. Abdullaev, B. B. Baizakov and E.
N. Tsoy for many fruitful discussions. This work has been
supported by the research grant No. $\Phi2-\Phi\mathrm{A}-0-97004$
of the Uzbek Academy of Science. \\


\section*{References}

\begin{thebibliography}{99}

\bibitem{Scott}
A. Scott, Encyclopedia of Nonlinear Science (Routledge, New York,
2005)

\bibitem{Mewes}
M. -O. Mewes, M. R. Andrews, D. M. Kurn, D. S. Durfee, C. G.
Townsend and W. Ketterle, {\it Phys. Rev. Lett.} {\bf 78}, 582
(1997).

\bibitem{Andrews}
M.R. Andrews, C. G. Townsend, H. Miesner, D. S. Durfee, D. M. Kurn
and W. Ketterle, {\it Science} {\bf275}, 637 (1997).

\bibitem{Merimeche}
H. Merimeche, {\it J. Phys. B: At. Mol. Opt. Phys.} {\bf39}, 3723
(2006).

\bibitem{Balykin}
V. I. Balykin, V. S. Letokhov, Yu. B. Ovchinnikov and A. I.
Sidorov, {\it Phys. Rev. Lett.} {\bf60}, 2137 (1988).

\bibitem{Friedrich}
H. Friedrich, G. Jacoby and C. G. Meister, {\it Phys. Rev.} {\bf
A65}, 032902 (2002).

\bibitem{Shimizu}
F. Shimizu, {\it Phys. Rev. Lett.} $\mathbf{86}$, 987 (2001); H.
Oberst, Y. Tashiro, K. Shimizu and F. Shimizu, {\it Phys. Rev.}
{\bf A71}, 052901 (2005).

\bibitem{Fujita}
F. Shimizu and J. Fujita, {\it J. Phys. Soc. Jpn.} {\bf 71}, 5
(2002); D. Kouznetsov and H. Oberst, {\it Opt. Rev.} {\bf12}, 363
(2005).

\bibitem{Kouznetsov}
H. Oberst, D. Kouznetsov, K. Shimizu, J. Fujita and F. Shimizu,
{\it Phys. Rev. Lett.} {\bf94}, 013203 (2005); D. Kouznetsov and
H. Oberst, {\it Phys. Rev.} {\bf A72}, 013617 (2005).

\bibitem{Pasquini}
T. A. Pasquini, Y. Shin, C. Sanner, M. Saba, A. Schirotzek, D. E.
Pritchard and W. Ketterle, {\it Phys. Rev. Lett.} {\bf 93}, 223201
(2004).

\bibitem{Ott}
H. Ott, J. Fortagh, G. Schlotterbeck, A. Grossmann and C.
Zimmermann, {\it Phys. Rev. Lett.} {\bf 87}, 230401 (2001); W.
Hansel, P. Hommelhoff, T. W. Hansch and J. Reichel, {\it Nature}
{\bf 413}, 498 (2001); Y. Shin, C. Sanner, G.-B. Jo, T. A.
Pasquini, M. Saba, W. Ketterle, D. E. Pritchard, M. Vengalattore
and M. Prentiss, {\it Phys. Rev.} {\bf A72}, R021604 (2005).

\bibitem{Saif}
F. Saif and I. Rehman, {\it Phys. Rev.} {\bf A75}, 043610 (2007).

\bibitem{Pethick}
C. J. Pethick, H. Smith, {\it Bose-Einstein Condensation in Dilute
Gases}, (Cambridge University Press, 2002).

\bibitem{Pitaevskii}
L. V. Pitaevskii, S. Stringari, {\it Bose-Einstein Condensation},
(Clarendon Press, Oxford, 2003).

\bibitem{Zakharov}
V.E. Zakharov, S.V. Manakov, S.P. Novikov, L.P. Pitaevskii, {\it
Theory of solitons: the inverse scattering method}, (Plenum,
Consultants Bureau, New York, 1984).

\bibitem{Khaykovich}
L. Khaykovich, F. Schreck, G. Ferrari, T. Bourdel, J. Cubizolles,
L. D. Carr, Y. Castin and C. Salomon, {\it Science} {\bf 296},
1290-1293 (2002).

\bibitem{Ford}
J. Ford and G. Lunsford, {\it Phys. Rev.} {\bf A1}, 59 (1970).

\bibitem{Stoker}
J.J. Stoker, {\it Nonlinear Vibrations in Mechanical and
Electrical Systems}. (New York, 1950).

\bibitem{Bogoliubov}
N. N. Bogoliubov, Y. A. Mitropolsky, {\ it Asymptotic Methods in
the Theory of Non-Linear Oscillations}, (New York, 1961).

\bibitem{Naifeh}
A. H. Naifeh, {\it Intriduction to Perturbation Techniqees}.
(Wiley, New York, 1981).

\bibitem{Tsoy}
E. N. Tsoy and B. A. Umarov, {\it Phys. Lett.} {\bf A235}, 147-152
(1997).

\bibitem{Moon}
F.C. Moon, {\it Chaotic Vibrations. An Introduction for Applied
Scientists and Engineers}. (A Wiley-Interscience publication,
1987).

\bibitem{Chirikov}
B. V. Chirikov, {\it Phys. Rep.} {\bf52}, 263 (1979).

\bibitem{Lichtenberg}
A. J. Lichtenberg, M. A. Liberman, {\it Regular and Stochastic
Motion}. (Springer, Berlin, 1983).

\bibitem{Hanggi}
P. Hanggi, P. Talkner and M. Borkovec, {\it Rev. Mod. Phys.} {\bf
62}, 251 (1990).

\bibitem{Nesvizhevsky} V. V. Nesvizhevsky, et al., {\it
Nature} {\bf 415}, 297 (2002).




\end{thebibliography}
\end{document}